\documentclass[preprint,authoryear,3p,twocolumn]{elsarticle}
\usepackage{graphicx}

\usepackage{ulem}

 \usepackage{graphics}
 \usepackage{graphicx}
 \usepackage{epsfig}

\usepackage{amssymb}

\journal{New Astronomy}

\begin{document}

\begin{frontmatter}



\title{Application of Neural Networks to  the study of stellar model solutions}


 \author[label1,label2,label3]{F.J.G. Pinheiro \corref{cor1}}
 \cortext[cor1]{Corresponding author}
 \ead{fpinheiro@teor.fis.uc.pt}
 \author[label4]{T. Simas}
 \author[label1,label2,label5]{J. Fernandes}
 \author[label4]{R. Ribeiro}
 \address[label1]{Centro de F\'{\i}sica Computacional da Universidade de Coimbra, Rua Larga, 3004--516 Coimbra, Portugal}
 \address[label2]{Observat\'{o}rio Astron\'{o}mico da Universidade de Coimbra, Santa Clara, 3040--004 Coimbra, Portugal}
 \address[label3]{Lesia, Observatoire de Paris, 5, place Jules Janssen, 92195 Meudon, France}
 \address[label4]{UNINOVA/CA3, Universidade Nova de Lisboa, Quinta da Torre, 2829--516 Monte de Caparica, Portugal}
 \address[label5]{Departamento de Matem\'{a}tica da Universidade de Coimbra, 3001--454 Coimbra, Portugal}

\begin{abstract}
Artificial neural networks (ANN) have different applications in Astronomy, 
including data reduction and data mining.
 In this work we propose the use ANNs in the identification of stellar model solutions.
 We illustrate this method, by applying an ANN to the 0.8M$_\odot$ star CG Cyg B.  
 Our ANN was trained using 60,000 different 0.8M$_\odot$ stellar models. 
 With this approach we identify the models which reproduce CG Cyg B's position in the 
HR diagram. 
 We observe a correlation between the model's initial metal and helium abundance which, 
 in most cases,  does not agree with a helium to metal enrichment ratio $\Delta$Y/$\Delta$Z=2.
   Moreover, we identify a correlation between the model's initial helium/metal abundance
 and both its age and mixing-length parameter. 
   Additionally, every model found has a mixing-length parameter below 1.3. This means that 
 CG Cyg B's mixing-length parameter is clearly smaller than the solar one. 
 From this study we conclude that ANNs are well suited to deal with the degeneracy 
  of model solutions of solar type stars.

\end{abstract}

\begin{keyword}
stars: evolution  \sep stars: fundamental parameters  \sep stars: interiors \sep stars: individual (CG Cyg B)
\end{keyword}

\end{frontmatter}

\section{Introduction}

 The determination of stellar masses and ages is vital for different areas of Astronomy 
\citep{catelan07} such as planetary formation \citep{johnson07}. 
 These parameters can be derived comparing the position of a star in the HR diagram (HRD) 
against model predictions  
\citep{lastennet02}. This is known as an HRD analysis. 
 Frequently, the computation of stellar evolutionary models involves the use parameters 
for which we do not have strong observational constraints. 
 Some of these are used to describe mechanisms, such as convection and diffusion, which 
are insufficiently known  
\citep{cassisi05}.
 As a consequence, we tend to have more modelling parameters than observational constrains, 
which results in a degeneracy of model solutions. 
 In order to reduce this problem, mass and age determinations are currently made using 
isochrones computed assuming solar scaled values for the helium abundance and convection 
parameters.  Yet this may lead to wrong results.

 An artificial neural network is composed of a collection of artificial neurons 
organized in several different structures, denoted architectures. 
 Each neuron receives inputs, processes the inputs and delivers a single output. 
 A typical structure has three layers: input, intermediate (called hidden layer) and output. 
 Its capacity for analysing large amounts of data 
\citep{bishop95} and its ability for dealing with multidimensional problems 
\citep{haykin99} makes them valuable in different areas of Astronomy such as data 
eduction and data mining 
\citep{tagliaferri03}. 
 Indeed, ANNs have been used to perform an automated classification of stellar spectra 
and determine global stellar parameters (luminosity L, effective temperature T$_{eff}$ and 
metal abundance [M/H])from low resolution spectra 
\citep{bailer00}.

 In this work we propose a new application of ANNs, the identification of stellar 
modelling parameters age, helium abundance (Y), metalicity (Z) and the mixing-length 
parameter $\alpha$ defined by 
\citet{bohm58} of solar type stars, by taking into account their position in the HR diagram. 
 This allows to analyse the degeneracy of stellar model solutions and evaluate its impact 
on the determination of stellar ages. 
  We conclude by presenting an application to the star CG Cyg B.

\section{The Artificial Neural Network}

\subsection{Overview and Motivation}

 The challenge in this work is to perform a  
multidimensional curve fitting for an astrophysical relationship between four 
variables, given a set of temperature and luminosity for a fixed mass 
\citep{cunha03}.

 Since the relation between the four variables is not known a priori, the multidimensional 
curve fitting is done using the inverse relation (four variables: Y, Z, $\alpha$ 
and age) to two variables (L \& T$_{eff}$). 
 Then a complete mapping is done by simulating the trained Neural network (NN) against all 
possible values of the four variables 
required for our case study.
 Yet why using a NN to tackle this regression problem?
 First of all, albeit having a large amount of input/output data is available we are 
not sure how to relate it. 
 Despite this problem appears to have overwhelming complexity, there is clearly 
a solution. 
 Indeed, it is relatively easy to create a number of examples of the correct behaviour. \\
 The objective of using a neural network (NN) model is to emulate a biological neural network 
\citep{haykin99,bishop95,mitchell97,duda01}.

 NN are composed of a collection of artificial neurons organized in several different structures, 
denoted architectures. Each neuron receives inputs, processes the inputs and delivers a single output. 
 A typical structure has three layers: input, intermediate (called hidden layer) and output. 
 There are several flavours of artificial neural networks: feed-forward, recursive, radial basis 
and many more \citep{haykin99}. 
 Back-propagation is a simple and successful training procedure to feed-forward NN and very useful 
for curve fitting \citep{duda01}. It is a supervised learning algorithm that uses the delta rule to 
minimize the error between the pattern and the classification at the output layer by back-propagating 
this error layer by layer actualizing the fitting parameters, the node weights. We have chosen this 
paradigm and in particular back-propagation NN because of its ability to deal with complex, non-linear 
and parallel computation \citep{haykin99}. An interesting reference to supporting tools for developing 
applications of NN is shown by \citet{demuth08}.
 Neural networks display remarkable capabilities to derive meaning from complicated or imprecise data 
and can be used to extract patterns and detect trends that are too complex to be noticed by either 
humans or other computer techniques \citep{stergiou96}.    
A trained neural network is similar to an 
"expert" analysing known information. Another interesting characteristic of NN is its capability for 
handling multidimensionality, which is related with three factors, dimensions, discovery and time 
\citep{bishop95,duda01}. 
 The main advantages of NN over other technologies are based in their following characteristics 
\citep{stergiou96}:\\
\indent 1. Adaptive learning: An ability to learn how to do tasks based on the data given for training or 
initial experience.\\
\indent 2. Self-Organisation: An NN can create its own organisation or representation of the information it 
receives during learning time.\\
\indent 3. Real Time Operation: NN computations may be carried out in parallel, and special hardware devices 
are being designed and manufactured which take advantage of this capability.\\
\indent 4. Fault Tolerance via Redundant Information Coding: Partial destruction of a network leads to the 
corresponding degradation of performance. However, some network capabilities may be retained even 
with major network damage”.\\

 For all the characteristics and capabilities of NN just described, it seems rather appropriate and 
interesting for addressing astronomy problems such as the one discussed here. Furthermore, NN are 
being recognized as a useful and flexible technology within Astrophysics domain 
\citep{tagliaferri03,andreon00}.
 In summary, the motivation for using a Neural Network (NN) is its modelling ability 
and versatility to handle classification problems, in large data sets with uncertain information, which 
seems rather suitable for studying degeneracy’s in stars. 

 Considering the need to demonstrate the capabilities of our devised NN approach for 
constraining stellar modelling parameters and detecting degeneracies of stellar model solutions,
we used a relatively contained case study as  proof-of-\hbox{-con}cept for our model.  
 The case study is based on data previously computed by \citet{fernandes03}, 
which comprises a regression of 60,000 M=0.8M$_\odot$ stellar models, with six 
variables: 4 outputs ($\alpha$, Y, Z \& age) and 2 inputs (T$_{eff}$ \& L). 
The stellar evolutionary models required for this analysis were computed using the CESAM 
stellar evolutionary code
 \citep{morel97}. 
 In these computations we used the same physical ingredients adopted by
 \citet{cunha03}
in the modelling of HR 1217.

\subsection{The Neural Network model}

 The NN model is based on the knowledge that Stellar masses and ages can be derived comparing 
the position of a star in the HR Diagram against the predictions of evolutionary models 
\citep{fernandes03}.
 This is known as an HRD analysis \citep{fernandes03,demuth08,tagliaferri03}. 
 In order to describe the stellar interiors we frequently use several parameters, such 
as the mixing-length parameter, whose values are uncertain. 
 This leaves us with an open problem. 
 Consequently, we find several combinations of modelling 
parameters which are able to reproduce the effective temperature and luminosity of a given star, 
which does not allow to accurately derive its mass and age. 
 As mentioned before we limited our computations to 0.8M$_\odot$ stellar models. 
 This study can be particularly useful for choosing the best Hertzprung-Russel Diagram (HRD) 
regions and infer stellar parameters from the knowledge of luminosity and effective temperature. 
 It is well known that this inverse problem has no unique solution \citep{fernandes03}.

 In the classical literature on Neural Networks (NN) there are many types of architectures and 
reasoning schemes that can used to model the problem at hand \citep{haykin99,bishop95,mitchell97}.
 However, from the set of possible architectures for artificial neural networks (Feed-Forward, 
Feed-Back, Network Layers, Perceptrons), we chose the Feed-Forward, where information is 
constantly ”fed forward” from one layer to the next, without loops, associating inputs with outputs. 
 The reason for the choice of using a back-propagation algorithm is due to its capability of 
working with large amounts of data and also its efficient tuning ability 
\citep{haykin99,andreon00,demuth08}.
 Considering that Back-Propagation Neural Networks learn by example, we used a set of 60\% of 
total input elements, 20\% for checking the data set, and 20\% for validation. 
 To train the NN we used the Levenberg-Marquardt algorithm \citep{demuth08},
which is a faster algorithm than the delta rule, to optimize the minimization of the error. 
 As transfer functions for the neurons, we used a sigmoid function in the first hidden layer 
and a linear function on the second layer (output). 

 The  $(Y,Z,\alpha,age)$=$f(T_{eff},L)$ regression function, is the result of the trained neural 
network, feed-forward with back-propagation for the error propagation. 
 Since the objective is to identify/analyse stellar model degeneracies, we address the problem by 
finding the inverse relation, for a given temperature and luminosity, given a fixed mass such as: 
$(T_{eff},L)=R(Y,Z,\alpha,Age)$, 
where R is one to many relation, because of the problem degeneracy. This relation is defined by 
using f to interpolate the values of a grid for all possible values of Y, Z, $\alpha$ and age, 
with a desired precision. 
 This inverse relation is the main novelty in our NN approach.

\subsection{Reasoning scheme}

\begin{figure}[ht]
 \centering
\includegraphics[width=6.5cm, height=6.5cm]{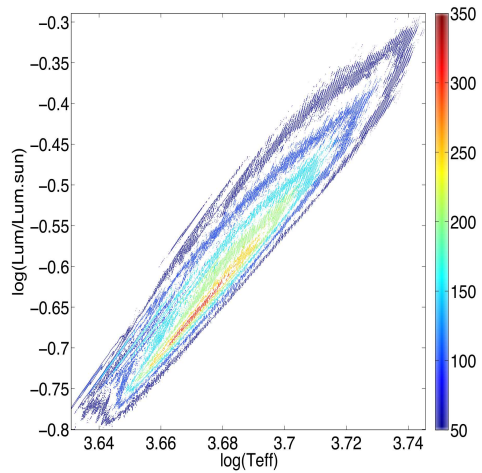}
\caption{Degeneracy, i.e. number of model solutions, for 0.8 M$_\odot$ stars  
for a bin in Teff and L of 1000. 
 The red areas correspond to regions of the HRD where we can find the highest  
degeneracy of 0.8 M$_\odot$ stellar model solutions.
} 		
\label{fig_bin}
\end{figure}

 Considering that our case study are 0.8M$_\odot$ population I stars, our demonstrator took into 
account the range of input variables given at Table 1. 
 This range of parameters is well suited to study population I stars located within the galactic 
disk  \citep{cunha03, fernandes03}.
 Moreover, we set the precision for each variable according to the binsize shown at Table 1. 
 This gives us about 10$^7$ grid points representing the possible universe of stars. By simulation 
of the Neural Network we get their respective effective temperature and luminosity, thus completing 
the interpolation. 
Figures such as Fig. \ref{fig_bin}, which in this case has a 1000 element bin in 
effective temperature and luminosity, helps us to identify the regions of the HRD where we should 
expect the highest degeneracy of model solutions.

\begin{table}[t]
  \centering
   \caption{Variable range and their precision}
 \begin{tabular}{ l c c c c } \hline
Variable   &  $Y$        &  $Z$        & $\alpha$   & $age (MYr)$     \\ \hline
 Lowest    &  $0.23$     & $0.008$     &  $1$       & $74$           \\
Highest    & $0.3$       & $0.03$      & $2$        & $10,000$       \\
 $\#$bins  & 10          & 100         & 10         & 1000            \\
 binsize   &   0.007     &  0.00022    & 0.1        &  992.6          \\
 MaxDev &   50\%      &  31.82\%    & 25\%       & 40.30\%        \\ \hline
  \end{tabular}
  \label{table1}
\end{table}

 The results obtained by training our network are: mean squares error of 7.03$\times10^{-7}$ 
which indicates a very low error between input and output values, and a regression value of 
0.9997, very close to 1, representing a close correlation between input and output values. 
 Using a 1000 element bin in effective temperature and luminosity (shown at Fig. \ref{fig_bin}) 
we notice that, for all bins, the standard deviation for the mixing-length parameter ranges 
from 0.05 to 0.25. 
 This allows to identify, for instance, 0.8M$_\odot$ stars with a mixing length smaller 
than the solar value.
 On its own turn, the standard deviation for the initial metallicity is less than 0.007. 
Likewise, the standard deviation for the initial helium abundance ranges between 0.005 and 0.03. 
 Assuming that the error on the parameter determination is three times this value, we can 
estimate Y with an uncertainty between 0.015 and 0.09. In the range of parameters 
evaluated here (0.23$\leq$Y$\leq$0.30), this corresponds to a relative error between 5 and 39\% 
(with an average relative error around 20\%). 
 This accuracy in helium determination is competitive in relation to methods based on grid 
interpolations
\citep{casagrande07}.

 This concludes the reasoning scheme description that will be implemented and validated for 
any data set given by different users. 
 At this stage, the input is a txt file with the input variables; the NN is trained in Matlab, 
using the model described; a txt file is the result of the training process (output). 
 This output txt file is then imported to the DB (MySQL) included in the demonstrative 
web\hbox{-in}terface already built. After querying the DB with SQL syntax, text files can be 
saved and used to plot graphics, such as shown in Fig. \ref{fig_bin}.
 After this stage we are sure the network is well trained and the whole decision process works.

\section{Application to CG Cyg B}

 In order to illustrate our ANN's capability for the identification 
of stellar model solutions we require a star with a mass similar to 
the one of the models used to train our tool.
 The eclipsing binary CG Cyg is located well within the solar 
vicinity \citep{popper98}. 
 This constrains its component's age, initial helium and metal 
abundances, well within the range of modelling parameters used 
to train our ANN. 
 Moreover, the fact that CG Cyg's lower mass component is a 
0.810$\pm$0.013M$_\odot$ star 
\citep{popper94}, makes it a suitable target for our ANN.

 Light curve analysis and I$_c$-band measurements allowed 
\citet{hillenbrand04} to estimate CG Cyg B's effective 
temperature (log(T$_{eff})$=3.674$\pm$0.006) and luminosity 
(log(L/L$_\odot$)= \hbox{-0.510}$\pm$0.030). 
 Taking into account these global stellar properties, our ANN 
identified 75412 sets of modelling parameters which reproduce, 
within the given uncertainties, CG Cyg B's position in the HRD. 
 These are shown in \hbox{Fig.\ref{fig_CG}a}. 
 Note that no model reproduces CG Cyg B's exact position in the HRD. 
 Yet, this is already expected since this star is lightly 
more massive than 0.8M$_\odot$. 

 The full range of possible modelling parameters is shown at 
Table \ref{tab_examples}.
 This shows that CG Cyg B admits almost all possible values for 
the initial helium and metal abundances (Figs. 2b and c). 
 However, as seen in \hbox{Fig.\ref{fig_CG}f}, these parameters are 
strongly correlated, presenting a linear correlation coefficient 
r=0.909 (corresponding to a false alarm probability smaller than 0.1\%).
 Indeed, a linear fit to the model solutions gives: 

\begin{equation}
  \label{eq_Z-Y}
 Z= -0.0275 + 0.1736 Y, 
\end{equation}

\noindent with a r.m.s.$\approx$0.0016.
 Figure \ref{fig_CG}f shows that, in most cases, the 
model's initial helium and metal abundances are not in agreement 
with what is expected from a helium to metal enrichment ratio 
$\Delta$Y/$\Delta$Z=2, which takes into account the solar helium 
to metal proportion \citep{casagrande07}.
 Knowledge on the star's metallicity would be valuable to test 
this enrichment ratio. 
 Figure \ref{fig_CG}h shows that there is some degree of 
correlation between the model's age and its chemical composition. 
 The best fit to the data corresponds to:

\begin{figure*}[htbp]
 \centering
\includegraphics[width=16cm, height=21cm]{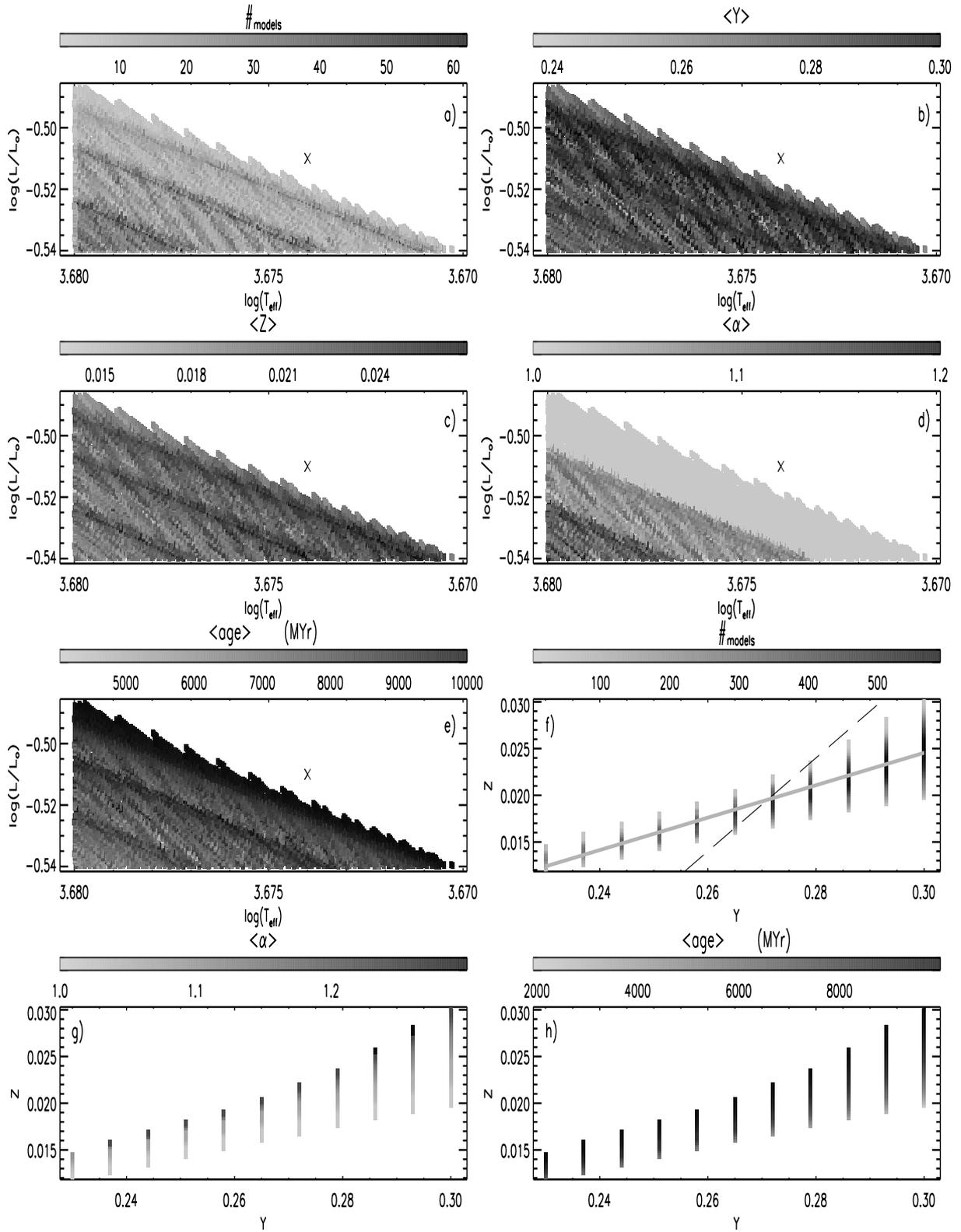}
\caption{
{\bf a)} Number of model solutions for CG Cyg B (100 bins in 
  log(T$_{eff}$) and log(L/L$_\odot$)). 
{\bf b)} Average initial helium abundance for the same bin size.
{\bf c)} Average metallicity for the same bin size.
{\bf d)} Average mixing length for the same bins.
{\bf e)} Average age for the same bin size.
 The crosses corresponds to CG Cyg B's exact position in the HRD. 
{\bf f)} Number of model solutions for CG Cyg B  
(0.07 bin size in Y and 0.022 in Z).
 The solid grey line corresponds to the best linear fit to the 
model's initial helium and metal abundances.
 The dashed line corresponds to a helium to metal enrichment ratio 
$\Delta$Y/$\Delta$Z=2.
{\bf g)} Average mixing-length parameter for different initial helium and 
 metal abundances for the same bins. 
{\bf h)} Average age for different initial helium and metal 
abundances of the same bin size.} 		\label{fig_CG}
\end{figure*}

\begin{table}[t]
  \centering
  \small
  \renewcommand{\tabcolsep}{3.0pt}
   \caption{Range of modelling parameters for CG Cyg B.}
  \begin{tabular}{ c  c  c   c  c  c  c  c } \hline
\multicolumn{2}{c}{Y} & \multicolumn{2}{c}{Z} & \multicolumn{2}{c}{$\alpha$} & \multicolumn{2}{c}{age (MYr)} \\    
  min     &    max    &     min   &    max    &       min      &      max    &   min     &  max    \\ \hline    
 0.23    & 0.30     & 0.01196   & 0.03   & 1.0            & 1.3           & 1850.754  & 10000.   \\ \hline
  \end{tabular}
  \label{tab_examples}
\end{table}

\begin{eqnarray}
  \label{eq_Ag-Y-Z}
age = 3.46\times10^4 -1.527\times10^5 Y     \nonumber \\
    +7.653\times10^5 Z ~ (MYr),
\end{eqnarray}

\noindent with a r.m.s.$\approx$902.
 In average, the model solutions seem to be older than the Sun 
(Fig.\ref{fig_CG}e). This is reinforced by the fact, in the 
bin size that we have selected, the age standard deviation of 
the best models (i.e. those that better reproduce CG Cyg's 
position in the HRD) is 1000MYr.
 Likewise,  $\alpha$ is correlated with the model's chemical 
composition (Figure \ref{fig_CG}g), with the best fit: 

\begin{equation}
  \label{eq_A-Y-Z}
 \alpha = 1.537 -3.462 Y +22.927 Z,   
\end{equation}

\noindent with a r.m.s.$\approx$0.055.
 Notice that every possible model has a mixing length smaller 
than 1.3  (Table \ref{tab_examples} and Figs. \ref{fig_CG}d and g). 
 This means the CG Cyg B's  $\alpha$ is clearly smaller than 
the solar value.
 \citet{lastennet02} found no 
isochrone that could fit CG Cyg B's position in the HR diagram, 
claiming that this star was far too cold. 
 Yet, they assumed a solar mixing-length parameter. 
 A lower mixing-length parameter (like the ones reported here) 
shifts the isochrones towards smaller effective temperatures. 

 Note that CG Cyg B is slightly more massive than the models used 
here. 
 Thus it is important to assess the impact that this mass difference 
can have on the parameter estimation.
 Studies of main sequence stars such as the components of the 
UV Psc binary
\citep{lastennet03} or sub giants like $\beta$ Hyd and evolutionary 
models \citep{fernandes03, pinheiro10} can provide us useful hints. 
In comparison with  0.80M$_\odot$ models, higher mass 
evolutionary tracks are shifted towards higher effective 
temperatures and luminosities.  
 Moreover 0.81M$_\odot$ stars evolve faster.
 Therefore, CG Cyg B's age should be slightly smaller than our models' 
predictions.
 The models computed by \citet{lastennet03} and \citet{pinheiro10} 
can be used to estimate the mixing-length's mass dependence. 
 Roughly speaking, the mixing-length's mass dependence 
 $\Delta$$\alpha$/$\Delta$M should be around -4,  
 i.e. our models overestimate CG Cyg B's mixing length parameter  
by a factor close to 0.04. 
 Thus reinforcing furthermore our conclusions regarding this parameter.
 The same models can be used to derive a similar $\Delta$Y/$\Delta$M ratio.
For instance, the overlap between Pinheiro \& Fernandes's evolutionary 
tracks of 0.90M$_\odot$ Y=0.29 and 0.92M$_\odot$ Y=0.25 stars indicates 
that our models overestimate CG Cyg B's helium abundance by a factor 
0.005. 
 Lastennet et al.'s models hint a similar result.
 Finally, the same reasoning applied to the metalicity gives a 
$\Delta$Z/$\Delta$M ratio around -0.1, i.e. an 0.001 overestimation 
of CG Cyg B's metalicity.
 This is close to what we obtain if we applied our helium overestimation 
to equation 1.

\section{Conclusions}

 The strong correlation observed between the initial helium and metal 
abundances 
of CG Cyg B's 0.80M$_\odot$ model solutions is similar to the one 
that can be seen in the analysis of UV Psc by
 \citet{lastennet03}.
 Yet, no numerical relationship is explicitly given there. 
 Moreover, we observe a correlation between the model's chemical 
composition (Y \& Z) and both their age and mixing-length parameter. 
 Finally notice that for all possible models, the later parameter 
($\alpha$) is smaller than the solar value.
 This could explain why for  binary stars like CG Cyg, isochrones 
fail to fit, at the same time, both component's position in the HRD. 

 Generally, we can constrain furthermore the modelling parameters of 
a given star by taking into account direct [Fe/X] observations and/or 
astreoseismic data. 
 In the particular case of CG Cyg, we can simultaneously analyse both 
components (which should have the same age, initial helium and metal 
abundance). 
 That leaves us with 5 unknown parameters (Y, Z, age, $\alpha_A$, 
$\alpha_B$) against 6 known global properties: M$_A$, T$_{eff A}$, 
L$_A$, M$_B$, T$_{eff B}$, L$_B$.
 Yet, this is not the scope of the present work. Indeed this work aims 
to show the adequacy of ANNs  for the identification of modelling 
parameters which reproduce known global stellar properties and, 
consequently, their usefulness for analysing the degeneracy of stellar 
model solutions.\\

 In this work we applied an ANN in order to perform a regression between the 
4 parameters that we are searching and the two observables. 
 The way our ANN is implemented ensures the reliability of this regression.
 Also notice that by treating the modelling parameters as free, we do not have 
the risk of suffering the consequences of assuming the wrong values.
 Additionally, once the ANN has been trained, our tool is able to identify 
the model solutions faster than approaches such as PSwarm 
\citep{fernandes11} 
which, at each iteration, require the computation of 
an additional stellar models. 
 This is particularly important in situations where a large number of 
stars have to be analysed.
 Likewise, our method's accuracy in helium determination is competitive in 
relation to methods based on grid interpolations 
\citep{casagrande07}
 On the other hand, the tool presented here is useful for the study of 
the degeneracy of stellar model solutions. 
 Generally, only a small amount of solutions are evaluated 
\citep[e.g.][]{fernandes03}, 
while here we have access to a large range of model solutions, 
allowing to analyse the degeneracy of model solutions problem as a whole.

 Having shown the suitability of our method, we plan to apply our ANN to a 
larger sample of stars. 
 This includes not only individual stars (for which their metal abundance 
may be known or not), binary stars and stellar populations, in 
particularly in clusters. This will allow studies on the chemical 
evolution of galaxies.
 Yet, in order to do so our Neural Network  has to be trained using 
other stellar masses.
The choice of the mass intervals needs to take into account 
the impact that this choice will have on the determination of stellar 
parameters.  
 That will be slightly different for different regions of the HRD. 
 In any case, a reasoning scheme similar to the one done here for 
CG Cyg B will be valuable.

 In the future we plan to implement 
models with different masses 
in the web-based tool, 
i.e. using the mass as an additional output parameter 
 The final result will be a tool that accepts txt files as inputs, 
trains and validates the ANN and produces outputs (both in text 
format and graphical form).  

 As a final remark we should remind that, like all works relying 
on the use of stellar evolutionary models, our ANN is limited by 
the set of models used to train it.
 This stresses the importance of using the right  physical ingredients 
and assumptions.

\section*{Acknowledgments}

 This work was supported by project PTDC/CTE-AST/66181/2006 
from Funda\c{c}\~{a}o para a Ci\^{e}ncia e a Tecnologia (FCT). 
F.~J.~G. Pinheiro also acknowledges FCT's grant SFRH/BPD/37491/2007. 
We wish to thank Bruno Pereira for his help in the development 
of the demonstrator web-based tool.

\bibliographystyle{elsarticle-harv}

\bibliography{arxiv_fjgp_NewA1203}

\end{document}